\documentstyle[12pt,preprint,aps,amsfonts,floats]{revtex} 
\tighten 
\newcommand{\be}{\begin{equation}}
\newcommand{\ee}{\end{equation}}
\newcommand{\bea}{\begin{eqnarray} \nonumber}
\newcommand{\eea}{\end{eqnarray}}
\newcommand{\beaa}{\begin{eqnarray}}
\newcommand{\eeaa}{\end{eqnarray}}
\newcommand{\ba}{\begin{array}}
\newcommand{\ea}{\end{array}}
\newcommand{\bit}{\begin{itemize}}
\newcommand{\eit}{\end{itemize}}
\newcommand{\ben}{\begin{enumerate}}
\newcommand{\een}{\end{enumerate}}
\newcommand{\bib}{\bibitem}

\def\lab{\label}
\def\lan{\langle}

\def\le{\left}

\def\noi{\noindent}
\def\non{\nonumber}

\def\ran{\rangle}
\def\rar{\rightarrow}

\def\ri{\right}

\def\al{\alpha}
\def\bt{\beta}

\def\de{\delta}

\def\ep{\epsilon}

\def\te{\theta}

\def\om{\omega}

\def\vec#1{{\bf #1}}

\newcommand{\mlab}[1]{\label{#1}}
\begin{document}
\title{New Results in the Physics of Neutrino Oscillations }

\author{{\bf Massimo Blasone}\thanks{e-mail: M.Blasone@ic.ac.uk. 
To appear in the Proceedings of the 
36th International School of Subnuclear 
Physics, Erice, 29 Aug - 7 Sep 1998.}}

\address{ Blackett Laboratory, Imperial College, Prince Consort
Road, \\ London SW7 2BZ, U.K.  
\\ and \\
Dipartimento di Fisica dell'Universit\`a di  Salerno 
\\  I-84100 Salerno, Italy} 
\maketitle 
\vspace{1cm}

\centerline{\bf Contents:}
\ben

\item
Introduction
\item
Mixing of Fermion Fields
\item
Neutrino Oscillations: the Exact Formula
\item
Berry Phase for Oscillating Neutrinos
\item
Discussion and Conclusions
\een


$$   $$



\noi{\bf  1. Introduction}

\vspace{0.2cm}

Neutrino mixing and oscillations\cite{Sol} 
are among the most interesting
topics of modern Particle Physics: at theoretical level, the
understanding of the origin of the small masses of neutrinos and of 
the mixing among generations is  still a puzzling problem\cite{Fri}. 
At experimental level, recent results\cite{Kam} seem
finally to confirm the existence of neutrino oscillations and
(consequently) of non vanishing masses for these particles.

However, since the pioneering work of Pontecorvo\cite{BP78}, 
who first pointed out 
the possibility of flavor oscillations for mixed massive
neutrinos, a 
careful  analysis of the structure of the
Hilbert space for mixed particles was not carried out successfully
\cite{Fuj,GKL92}.

This was achieved only recently\cite{Baop} and 
in the present paper I report about these results. 
I show that a study of the mixing transformations in 
Quantum Field Theory (QFT), reveals a rich non-perturbative structure of
the vacuum for the mixed fields.
This fact has phenomenological consequences on neutrino oscillations:
the oscillation formula turns out to have an additional 
oscillating piece and energy
dependent amplitudes, in contrast with the usual (quantum mechanical)
Pontecorvo formula, 
which is however recovered in the relativistic limit. 
I also show how the concept of a topological (Berry)
phase  naturally enters the physics of neutrino oscillations\cite{ber}.

%
\newpage

\noi{\bf  2. Mixing of Fermion Fields}

\vspace{0.2cm}

In order to discuss neutrino oscillations in QFT, I consider the
following Lagrangian for two Dirac fields $\nu_{e}$ and
$\nu_{\mu}$ (omit space-time dependence for brevity)
\begin{equation}\mlab{lag2}  
{\cal L} = {\bar \nu}_e
\left(i\not\partial -m_{e}\right)\nu_e + {\bar   
  \nu}_\mu\left(i\not\partial - m_{\mu}\right)\nu_\mu  
- \; m_{e \mu} \;\left({\bar \nu}_{e}  
\nu_{\mu} + {\bar \nu}_{\mu} \nu_{e}\right)    
\;,\end{equation}  
which is sufficient to describe the single particle evolution of a mixed
fermion\footnote{The inclusion of interaction terms in the Lagrangian
eq.(\ref{lag2}) does not alter the following discussion which is about
the transformations eq.(\ref{2.1}). For a discussion of 
three flavor mixing, see ref.\cite{Baop}.}. 
Mixing arise when the above Lagrangian is diagonalized by means
of the transformations
\bea
&&\nu_{e}(x) =\; \nu_{1}(x) \; \cos\te   \,+\, \nu_{2}(x) \; \sin\te \\
\mlab{2.1}
&&\nu_{\mu}(x) =- \nu_{1}(x) \; \sin\te  \, + \,\nu_{2}(x)\; 
\cos\te\;, 
\eea
where $\te$ is the mixing angle. 
$\nu_{e}$ and $\nu_{\mu}$ are neutrino fields 
with definite flavors. 
$\nu_{1}$ while $\nu_{2}$ are (free) neutrino 
fields with definite masses $m_{1}$ and $m_{2}$, respectively.
In terms of $\nu_{1}$ and $\nu_{2}$, the above 
Lagrangian reads
\begin{equation}\mlab{lag3}  
{\cal L} = {\bar \nu}_1\left(i\not\partial -m_{1}\right)\nu_1 
+ {\bar \nu}_2\left(i\not\partial - m_{2}\right)\nu_2   
\;.\end{equation}  
with $ m_{e} = m_{1}\cos^{2}\theta \,+\, m_{2} \sin^{2}\theta~$, 
$m_{\mu} = m_{1}\sin^{2}\theta \,+\, m_{2} \cos^{2}\theta~$,  
$m_{e\mu} =(m_{2}-m_{1})\sin\theta \cos\theta\,$.

The free fields $\nu_{1}(x)$ and $\nu_{2}(x)$ are written as 
\be\mlab{2.2}
\nu_{i}(x) = \frac{1}{\sqrt{V}} \sum_{{\bf k},r}
\le[u^{r}_{{\bf k},i}e^{-i\omega_{k,i}t}
\al^{r}_{{\bf k},i}\:+ v^{r}_{-{\bf k},i}e^{i\omega_{k,i}t}
\bt^{r\dag }_{-{\bf k},i} \ri]e^{i {\bf k}\cdot {\bf x}}, \;\;\;
 ~ i=1,2 \;. 
\ee
with $\om_{k,i}=\sqrt{\vec{k}^2+m_i^2}$. 
The $\al_{i}$ and the $\bt_{i}$ ( 
$ i=1,2 \,$), are defined with respect to 
the vacuum state  $|0\ran_{1,2}$:
$\al_{i}|0\ran_{1,2}= \bt_i|0\ran_{1,2}=0$.
The anticommutation relations are the usual ones.
The orthonormality and  completeness relations are:
\be 
u^{r\dag}_{{\bf k},i} u^{s}_{{\bf k},i} =  
v^{r\dag}_{{\bf k},i} v^{s}_{{\bf k},i} = \de_{rs} 
\;,\qquad  u^{r\dag}_{{\bf k},i} v^{s}_{-{\bf k},i} = 
v^{r\dag}_{-{\bf k},i} u^{s}_{{\bf k},i} = 0\;,\qquad
\sum_{r}(u^{r}_{{\bf k},i} u^{r\dag}_{{\bf k},i} + 
v^{r}_{-{\bf k},i} v^{r\dag}_{-{\bf k},i}) = {\Bbb I}\;. 
\ee

Our first step is the study of the 
generator of eqs.(\ref{2.1}) and of 
the underlying group theoretical structure. 

Eqs.(\ref{2.1}) can be put in the form\cite{Baop}:
\begin{eqnarray}\mlab{exnue1}  
\nu_{e}^{\alpha}(x)    
&=& G^{-1}_{\theta}(t)\, \nu_{1}^{\alpha}(x)\, G_{\theta}(t)  
= \frac{1}{\sqrt{V}} \sum_{{\bf k},r}  
  \left[ u^{r,\alpha}_{{\bf k},1}e^{-i\omega_{k,1}t} \alpha^{r}_{{\bf
k},e}(t) +   
v^{r,\alpha}_{-{\bf k},1}
e^{i\omega_{k,1}t}\beta^{r\dag}_{-{\bf k},e}(t)  
\right]  e^{i {\bf k}\cdot{\bf x}} ,
\\ \mlab{exnum1} 
\nu_{\mu}^{\alpha}(x)   
&=& G^{-1}_{\theta}(t)\, \nu_{2}^{\alpha}(x)\, G_{\theta}(t)  
=  \frac{1}{\sqrt{V}}\sum_{{\bf k},r}  
\left[ u^{r,\alpha}_{{\bf k},2}e^{-i\omega_{k,2}t} \alpha^{r}_{{\bf
k},\mu}(t) +   
v^{r,\alpha}_{-{\bf k},2}
e^{i\omega_{k,2}t}\beta^{r\dag}_{-{\bf k},\mu}(t)  
\right] e^{i {\bf k}\cdot{\bf x}} , 
\end{eqnarray}
where $\al=1,..,4$. The annihilation operators 
for the flavor fields are defined as (indices are suppressed): 
$\alpha_{e,\mu}(t)\equiv G^{-1}_{\theta}(t)\, 
\alpha_{1,2}\, G_{\theta}(t)$ and 
$\bt_{e,\mu}(t)\equiv G^{-1}_{\theta}(t)\, 
\bt_{1,2}\, G_{\theta}(t)$ .

\newpage

The generator  $G_{\theta}(t)$ is given by
\be\mlab{2.11}
G_\te(t) = exp\Big[\te\Big(S_{+}(t) - S_{-}(t)\Big)\Big]\;. 
\ee
%
%
\be\mlab{2.10}
S_{+}(t) \equiv  \int d^{3}\vec{x} \; \nu_{1}^{\dag}(x)
\nu_{2}(x) \;\;,\;\;\;\;\;
S_{-}(t) \equiv  \int d^{3}\vec{x} \; \nu_{2}^{\dag}(x)
\nu_{1}(x)\,, 
\ee
and is (at finite volume) an unitary operator:
$G^{-1}_{\theta}(t)=G_{-\te}(t)=G^{\dag}_\te(t)$, 
preserving the canonical anticommutation relations. 

Eqs.(\ref{exnue1}), (\ref{exnum1}) follow from  
$\frac{d^{2}}{d\theta^{2}}\nu^{\alpha}_{e}=-\nu^{\alpha}_{e}\;,\;\;\;
\frac{d^{2}}{d\theta^{2}}\nu^{\alpha}_{\mu}=-\nu^{\alpha}_{\mu}$ with 
the initial conditions 
$\nu^{\alpha}_{e}|_{\theta=0}=\nu^{\alpha}_{1}$,  
$\frac{d}{d\theta}\nu^{\alpha}_{e}|_{\theta=0}=\nu^{\alpha}_{2}$ and 
$\nu^{\alpha}_{\mu}|_{\theta=0}=\nu^{\alpha}_{2}$, 
$\frac{d}{d\theta}\nu^{\alpha}_{\mu}|_{\theta=0}=-\nu^{\alpha}_{1}$.

Note  that $G_\te$ is an element of  $SU(2)$. 
Indeed, if one introduces
\be\mlab{2.12}
S_{3} \equiv \frac{1}{2} \int d^{3}\vec{x} 
\le(\nu_{1}^{\dag}(x)\nu_{1}(x) - 
\nu_{2}^{\dag}(x)\nu_{2}(x)\ri)\;, 
\ee
and the total charge (Casimir operator),
\be\mlab{2.13} 
S_{0} \equiv \frac{1}{2} \int d^{3}\vec{x} 
\le(\nu_{1}^{\dag}(x)\nu_{1}(x) + 
\nu_{2}^{\dag}(x)\nu_{2}(x)\ri)\;, 
\ee
then the $su(2)$ algebra is closed:
\be\mlab{2.14}
[S_{+} , S_{-}]=2S_{3} \;\;\;,\;\;\; [S_{3} , S_{\pm} ] = \pm S_{\pm}
\;\;\;,\;\;\;[S_{0} , S_{3}]= [S_{0} , S_{\pm} ] = 0\;. 
\ee

The crucial point about the above generator is that it does not leave
invariant the vacuum $|0 \ran_{1,2}$. Its action on it results in a new
state, 
\be\mlab{2.22}
|0 (t)\ran_{e,\mu} \equiv G^{-1}_{\theta}(t)\; |0 \ran_{1,2}\;, 
\ee
which is the flavor vacuum, i.e. the vacuum for the flavor fields.
Let us define $|0\ran_{e,\mu}\equiv |0(0)\ran_{e,\mu}$ and 
compute $_{1,2}\lan0|0\ran_{e,\mu}$. By writing
$|0\ran_{e,\mu}=\prod_{{\bf k}}|0\ran_{e,\mu}^{\bf k}$, 
we obtain\cite{Baop}:
\be
_{1,2}\lan 0|0\ran_{e,\mu}  =
\prod_{{\bf k}}\le(1- \sin^{2}\te\;|V_{{\bf k}}|^2\ri)^{2}
=e^{\sum_{{\bf k}}ln\;\le(1- \sin^{2}\te\;|V_{{\bf k}}|^2\ri)^{2} },
\ee
where the function $V_{{\bf k}}$ is defined in 
eq.(\ref{2.39}). Note that
 $|V_{\bf k}|^2$ depends on ${\bf k}$ 
only through its modulus, takes values
in the interval $[0,\frac{1}{2}[$ and $|V_{{\bf k}}|^2 \rar 0$ 
when $|{\bf k}| \rar \infty$ (see Fig.1).

By using the customary
continuous limit relation $\sum_{{\bf k}}\;\rar \; 
\frac{V}{(2\pi)^{3}}\int d^{3}{\bf k}$, in the infinite volume limit we  
obtain
\be\mlab{2.34}
\lim_{V \rar \infty}\; _{1,2}\lan0|0\ran_{e,\mu} =
\lim_{V \rar \infty}\; e^{\frac{V}{(2\pi)^{3}}\int d^{3}{\bf k} 
\;ln\;\le(1- \sin^{2}\te\;|V_{{\bf k}}|^2\ri) }= 0 
\ee
for any value of  the parameters $\te$, $m_{1}$ and $m_{2}$.

Eq.(\ref{2.34}) expresses the unitary inequivalence  
of the flavor and the mass representations 
and shows the non-trivial nature of the mixing 
transformations (\ref{2.1}). In other words, the mixing 
transformations induce a structure in the 
flavor vacuum which indeed turns out to be an $SU(2)$ generalized 
coherent state\cite{Per} (cf. eqs.(\ref{2.22}) and (\ref{2.11})). 

Of course, the orthogonality between 
$|0\ran_{e,\mu}$ and $|0\ran_{1,2}$ disappears when
$\te =0$ and/or  $m_{1} = m_{2}$ (in this case $V_{{\bf 
k}}=0$ for any ${\bf k}$ and also no mixing occurs in the Lagrangian
(\ref{lag2})). 

\newpage

Let us now look at the  explicit form for the  flavor annihilation 
operators. 
Without loss of generality, we can choose the reference frame
such that ${\bf k}=(0,0,|\vec{k}|)$. 
In this case the annihilation operators have the simple
form\footnote{The flavor operators of eq.(\ref{2.36}) 
do annihilate the flavor vacuum. For example:
$\al^{r}_{{\bf k},e}(t)|0 (t)\ran_{e,\mu}=
 G^{-1}_{\theta}(t)\, 
\alpha^{r}_{{\bf k},1} G_{\theta}(t)G^{-1}_{\theta}(t)
|0 \ran_{1,2} =0\,.$}:
\bea
&&\al^{r}_{{\bf k},e}(t)=\cos\te\;\al^{r}_{{\bf 
k},1}\;+\;\sin\te\;\le(
U_{{\bf k}}^{*}(t)\; \al^{r}_{{\bf k},2}\;+\;\ep^{r}\;
V_{{\bf k}}(t)\; \bt^{r\dag}_{-{\bf k},2}\ri) \\
\mlab{2.36}
&&\al^{r}_{{\bf k},\mu}(t)=\cos\te\;\al^{r}_{{\bf 
k},2}\;-\;\sin\te\;\le(
U_{{\bf k}}(t)\; \al^{r}_{{\bf k},1}\;-\;\ep^{r}\;
V_{{\bf k}}(t)\; \bt^{r\dag}_{-{\bf k},1}\ri)  \\
\non
&&\bt^{r}_{-{\bf k},e}(t)=\cos\te\;\bt^{r}_{-{\bf 
k},1}\;+\;\sin\te\;\le(
U_{{\bf k}}^{*}(t)\; \bt^{r}_{-{\bf k},2}\;-\;\ep^{r}\;
V_{{\bf k}}(t)\; \al^{r\dag}_{{\bf k},2}\ri) \\
\non
&&\bt^{r}_{-{\bf k},\mu}(t)=\cos\te\;\bt^{r}_{-{\bf 
k},2}\;-\;\sin\te\;\le(
U_{{\bf k}}(t)\; \bt^{r}_{-{\bf k},1}\;+\;\ep^{r}\;
V_{{\bf k}}(t)\; \al^{r\dag}_{{\bf k},1}\ri) 
\eea
where $\ep^{r}=(-1)^{r}$. Also,  $V_{{\bf k}}(t)=|V_{{\bf 
k}}|\;e^{i(\om_{k,2}+\om_{k,1})t}$ and $U_{{\bf k}}(t)=|U_{{\bf
k}}|\;e^{i(\om_{k,2}-\om_{k,1})t} $, with
\bea
&&|U_{{\bf k}}|\equiv u^{r\dag}_{{\bf k},2}u^{r}_{{\bf k},1}=
v^{r\dag}_{-{\bf k},1}v^{r}_{-{\bf k},2} \\
\mlab{2.37}
&&|V_{{\bf k}}|\equiv \ep^{r}\;u^{r\dag}_{{\bf 
k},1}v^{r}_{-{\bf k},2}=
-\ep^{r}\;u^{r\dag}_{{\bf k},2}v^{r}_{-{\bf k},1}\,. 
\eea
Explicitly,
\bea
&&|U_{{\bf 
k}}|=\le(\frac{\om_{k,1}+m_{1}}{2\om_{k,1}}\ri)^{\frac{1}{2}}
\le(\frac{\om_{k,2}+m_{2}}{2\om_{k,2}}\ri)^{\frac{1}{2}}
\le(1+\frac{|\vec{k}|^{2}}{(\om_{k,1}+m_{1})(\om_{k,2}+m_{2})}\ri)\\
\mlab{2.39}
&&|V_{{\bf k}}|=\le(\frac{\om_{k,1}+m_{1}}{2\om_{k,1}}
\ri)^{\frac{1}{2}}
\le(\frac{\om_{k,2}+m_{2}}{2\om_{k,2}}\ri)^{\frac{1}{2}}
\le(\frac{\vec{k}}{(\om_{k,2}+m_{2})}-
\frac{\vec{k}}{(\om_{k,1}+m_{1})}\ri) 
\eea
\be\mlab{2.40}
|U_{{\bf k}}|^{2}+|V_{{\bf k}}|^{2}=1 
\ee
We thus see that, at the level of annihilation operators, the structure
of the mixing transformation is that of a Bogoliubov
transformation nested into a rotation. The functions $U_{{\bf k}}$ and 
$V_{{\bf k}}$ play indeed the role of Bogoliubov coefficients.

\begin{figure}[t]
\setlength{\unitlength}{1mm}
\vspace*{80mm}
\includegraphics{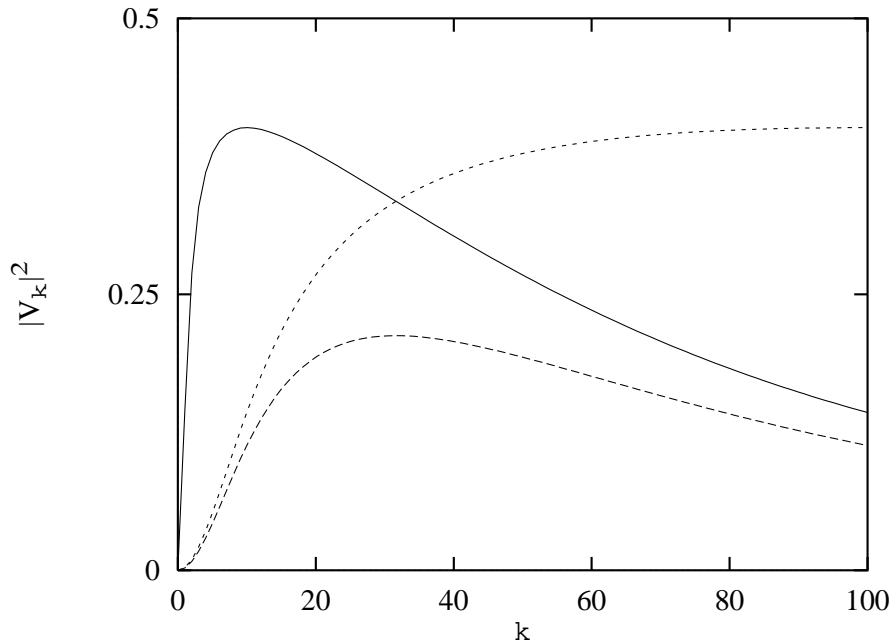}
\caption{The fermion condensation density $|V_{\bf k}|^2$ in 
function of  ${|\bf k|}$ and for sample values 
of the parameters $m_1$ and $m_2$. 
Solid line: $m_1=1 \;,\; m_2=100\;$; 
Long-dashed line: $m_1=10 \;,\; m_2=100\;$;
Short-dashed line: $m_1=10 \;,\; m_2=1000$.} 

\vspace{.1cm}

\hrule
\end{figure} 

It is also possible to exhibit the full explicit expression of 
$|0\ran_{e,\mu}^{{\bf k}}$ (at time $t=0$ and for
${\bf k}=(0,0,|{\bf k}|)$):  
\begin{eqnarray}\mlab{2.41}
&&|0\ran_{e,\mu}^{{\bf k}}= \prod_{r} \le[(1-\sin^{2}\te\;|V_{{\bf 
k}}|^{2}) 
-\ep^{r}\sin\te\;\cos\te\; |V_{{\bf k}}| 
\; (\al^{r\dag}_{{\bf k},1}\bt^{r\dag}_{-{\bf k},2}\;+
\;\al^{r\dag}_{{\bf k},2} \bt^{r\dag}_{-{\bf k},1})+\ri. \\
\non
&&\le.
+\;\ep^{r}\sin^{2}\te \;|V_{{\bf k}}| \;|U_{{\bf k}}| \le(
\al^{r\dag}_{{\bf k},1}\bt^{r\dag}_{-{\bf k},1}\; - \;
\al^{r\dag}_{{\bf k},2}\bt^{r\dag}_{-{\bf k},2} \ri) +
\sin^{2}\te \; |V_{{\bf k}}|^{2} \;\al^{r\dag}_{{\bf 
k},1}\bt^{r\dag}_{-{\bf k},2}
\al^{r\dag}_{{\bf k},2}\bt^{r\dag}_{-{\bf k},1}\ri]|0\ran_{1,2}
\end{eqnarray}
We see that the expression of the 
flavor vacuum $|0\ran_{e,\mu}$ 
involves four different 
particle-antiparticle ``couples''.
It is interesting to compare
$|0\ran_{e,\mu}$ with the 
BCS superconducting ground state\cite{BCS}, 
which involves only one kind of couple and is generated by a 
Bogoliubov transformation.

Finally, the condensation density is given by
\be\mlab{2.41b}
\;_{e,\mu}\lan 0| \al_{{\bf k},i}^{r \dag} \al^r_{{\bf k},i} 
|0\ran_{e,\mu}\,= 
\;_{e,\mu}\lan 0| \bt_{{\bf k},i}^{r \dag} \bt^r_{{\bf k},i} 
|0\ran_{e,\mu}\,=\,
\sin^{2}\te\; |V_{{\bf k}}|^{2} \;, \qquad i=1,2\,.
\ee
The function $|V_{{\bf k}}|^{2}$ has a maximum  
at $|{\bf k}|=\sqrt{m_1 m_2}$ (see Fig.1).

\newpage

%

\noi{\bf 3. Neutrino Oscillations: the Exact Formula}

\vspace{0.2cm}

Equipped with the results of the previous Section, we
can now study the time evolution of a flavor state, i.e. neutrino
oscillations. 
We know from  eq.(\ref{2.34}) that we have to work in the 
Hilbert space built on $|0\rangle_{e,\mu}$, since this is the space for
the flavor fields. 
 
At time $t=0$, the vacuum state is $|0\rangle_{e,\mu}$  
and the one electron neutrino state is ( for 
${\bf k}=(0,0,|{\bf k}|)$):   
\begin{equation}\mlab{h1}  
|\nu_e \rangle \equiv 
\alpha_{{\bf k},e}^{r \dag}|0\rangle_{e,\mu} = \left[   
\cos\theta\,\alpha_{{\bf k},1}^{r \dag} +  
|U_{\bf k}|\; \sin\theta\;\alpha_{{\bf k},2}^{r \dag} + 
\epsilon^r \; |V_{\bf k}| \,\sin\theta \;  
\alpha_{{\bf k},2}^{r \dag}\alpha_{{\bf k},1}^{r \dag}
\beta_{-{\bf k},1}^{r \dag}  \right] |0\rangle_{1,2} \,.  
\end{equation}  
In this state a multiparticle component is present,   
disappearing in the relativistic limit $|{\bf k}|\gg \sqrt{m_1m_2}\,$:
in this    
limit the (quantum-mechanical) Pontecorvo state is recovered.  

If we now assume that the neutrino state at time $t$ is given by
$|\nu_e (t)\rangle = e^{-iH t} |\nu_e\rangle$, we see that it is not
possible to compare directly 
this state with the one at time $t=0$ given in
eq.(\ref{h1}). The reason is that the flavor vacuum   
$|0\rangle_{e,\mu}$ is not eigenstate of the free Hamiltonian $H$ 
and it ``rotates'' under the action of the time  
evolution generator: one indeed finds 
$\lim_{V \rightarrow \infty}\;_{e,\mu}\langle 0\;|\;0(t)\rangle_{e,\mu} =  
0$. Thus at different times we have unitarily inequivalent flavor  
vacua and this implies that  we cannot directly compare flavor
states at different times. 
A way to circumvent this problem is to study the propagators for the
mixed fields $\nu_e$, $\nu_\mu$\cite{Bpre2}.

The crucial point is about how to compute these propagators: if one
(naively) uses the vacuum $|0\rangle_{1,2}$, one gets an inconsistent 
result (cf. eq.(\ref{pee2})). Let us show this by considering 
the Feynman propagator for electron neutrinos,
\begin{equation}\mlab{matrix}  
S_{ee}(x,y)\,=\,_{1,2}\lan 0 |
T \left[\nu_{e}(x) \bar{\nu}_{e}(y)\right] |
0\ran_{1,2}   
\end{equation}  
where $T$ denotes time ordering. Use of (\ref{2.1}) gives  
$S_{ee}$ in momentum representation as  
\begin{equation}\mlab{pert1}  
S_{ee}(k_0,{\bf k})=   
\cos^2\theta\; \frac{\not \! k + m_1}{k^2 - m_{1}^{2} + i\delta} \; +   
\;\sin^2\theta\; \frac{\not \! k + m_2}{k^2 - m_{2}^{2} + i\delta}  \; ,
\end{equation}  
which is just the weighted sum of the two propagators for the free   
fields $\nu_1$ and $\nu_2$. It coincides with the   
Feynman propagator obtained by resumming (to all orders)   
the perturbative series  
\begin{equation}\mlab{pert2} 
S_{ee} = S_e\left(1\,+\,m_{e\mu}^2 \,S_\mu S_e \,+\, m_{e\mu}^4 \,S_\mu   
S_e S_\mu S_e \,+\, ...\right) = 
S_e\left(1 - m_{e\mu}^2 \,S_\mu S_e \right)^{-1}\,,  
\end{equation} 
where the ``bare'' propagators are defined as $S_{e/\mu}=(\not \! k -   
m_{e/\mu} + i\delta)^{-1}$.    

The transition amplitude for an electron neutrino created   
by $\alpha_{{\bf k},e}^{r \dag}$ at time $t=0$ 
going into the same  
particle at time $t$, is given by 
\begin{equation}\mlab{exnue4}  
{\cal P}^r_{ee}({\bf  k},t)= i   
u^{r \dag}_{{\bf k},1}e^{i\omega_{k,1}t}\,  
S^>_{ee}({\bf  k},t)\,\gamma^0 u^r_{{\bf k},1}\, ,  
\end{equation}  
where the spinors $u_1$ and  $v_1$ form the basis in which the field
$\nu_e$ is expanded (cf. eq.(\ref{exnue1})).
Here, $S^>_{ee}$   
denotes the unordered Green's function (or Wightman function):
$i S^{>}_{ee}(t,{\bf x};0,{\bf y}) =   
{}_{1,2}\langle0|\nu_{e}(t,{\bf x}) \;  
\bar{\nu}_{e}(0,{\bf y})  |0\rangle_{1,2}$ .   
The explicit expression for $S^{>}_{ee}({\bf k},t)$ is  
\begin{equation}\mlab{gre2}  
S^{>}_{ee}({\bf k},t) =  
 -i\sum_{r}   
 \left( \cos^2\!\theta\;  
 e^{-i\omega_{k,1} t}\; u^{r}_{{\bf k},1}\;  
  \bar{u}^{r}_{{\bf k},1} \;   
 + \sin^2\!\theta\;  
 e^{-i\omega_{k,2} t} \;u^{r}_{{\bf k},2} \;  
 \bar{u}^{r}_{{\bf k},2}\;  
 \right)   
\;.\end{equation}  
The amplitude eq.(\ref{exnue4}) 
is independent of the spin orientation and given by  
\begin{equation}\mlab{pee1}  
{\cal P}_{ee}({\bf k},t)  
  =\cos^2\!\theta\,  
 + \sin^2\!\theta\,|U_{\bf k}|^{2}\, e^{-i(\omega_{k,2}-\omega_{k,1}) t}  
\;.\end{equation}  

For different masses and for ${\bf k}\neq 0\,$, 
$|U_{\bf k}|^2$ is always $<1$ (cf. eq.(\ref{2.39})).
Notice also that $|U_{\bf k}|^2\rightarrow 1$ 
in the relativistic limit $ |{\bf k}|\gg \sqrt{m_1 m_2}\;$: only in this
limit the squared modulus of ${\cal P}_{ee}({\bf k},t)$ 
does reproduce the Pontecorvo oscillation formula. 
  
Of course, it should be $\lim_{t\rightarrow 0^+} {\cal  
P}_{ee}(t)=1$. Instead, one obtains the unacceptable result 
\begin{equation}\mlab{pee2}  
{\cal P}_{ee}({\bf k},0^+)=  
\cos^2\!\theta + \sin^2\!\theta\, |U_{\bf k}|^2 < 1  
\;.\end{equation}  
This means that the {\em choice of the state\/} $|0\rangle_{1,2}$ in  
eq.(\ref{matrix}) and in the computation of the Wightman function  
is {\em not\/}  the correct one.  
The reason is that, as previously shown,
  $|0\rangle_{1,2}$ is {\em not} the vacuum state
for the flavor fields \cite{Baop}.
  
We are then led to define the propagators on the 
flavor vacuum $|0\rangle_{e,\mu}$. Considering again the  propagator for
the electron neutrinos, we have
\begin{equation}\mlab{matrix2}
G_{ee}(x,y)
\equiv  \,_{e,\mu}\langle 0(y_0) |   
T \left[\nu_{e}(x) \bar{\nu}_{e}(y)\right]  
|0(y_0) \rangle_{e,\mu}  \, .
\end{equation}
Notice that here the time argument $y_0$ (or, equally well, $x_0$)   
of the flavor ground state, is chosen to be equal on both sides   
of the expectation value. 
This is necessary since, as already observed,
transition matrix elements of the type  
${}_{e,\mu}\langle0|\alpha_e\,\exp\left[ -i H t\right]\,  
\alpha^{\dag}_e | 0\rangle_{e,\mu}$, do not represent    
physical transition amplitudes: they actually vanish (in the infinite
volume limit) due to the unitary inequivalence of flavor vacua at
different times\cite{Bpre2}. 
Therefore the comparison of states at different times  
necessitates a {\em parallel transport\/} of these states to  
a common point of reference. The definition (\ref{matrix2}) includes  
this concept of parallel transport, which is a sort of
``gauge fixing'': a geometric structure associated with  the 
simple dynamical system of eq.(\ref{lag2}) is thus uncovered. 
  
In mixed (${\bf k}, t$) representation, we have
(for ${\bf k}=(0,0,|{\bf k}|)$):  
\begin{eqnarray} \nonumber  
G_{ee}(k_0,{\bf k})&=&S_{ee}(k_0,{\bf k})\; + \;  2\pi\,i\, 
\sin^2\theta \Big[|V_{{\bf k}}|^2 
\; (\not \! k + m_2)\;\delta(k^2 - m_2^2)     
\\  \mlab{diff}  
&&  \;-\; |U_{{\bf k}}| |V_{{\bf k}}| 
\;\sum_{r} \left(\epsilon^r u^{r}_{{\bf k}   
,2}\;\bar{v}^{r}_{-{\bf k},2} \; \delta(k_0 - \omega_2 )\; + \;\epsilon^r   
v^{r}_{-{\bf k},2} \;\bar{u}^{r}_{{\bf k} ,2}\;\delta(k_0 + \omega_2 )   
\right)\Big]  \, ,
\end{eqnarray}  
Comparison of eq.(\ref{diff}) with eq.(\ref{pert1}) 
shows that the difference between the   
full and the perturbative propagators is in the imaginary part.  
  
I define the Wightman functions for an electron neutrino as   
$i G^{>}_{ee}(t,{\bf x};0,{\bf y}) =   
{}_{e,\mu}\langle0|\nu_{e}(t,{\bf x}) \;  
\bar{\nu}_{e}(0,{\bf y})  |0\rangle_{e,\mu}$, 
and $i G^{>}_{\mu e} (t,{\bf x};0,{\bf y}) =   
{}_{e,\mu}\langle0|   
\nu_{\mu}(t,{\bf x}) \; \bar{\nu}_{e}(0,{\bf y})  
|0\rangle_{e,\mu}$. 
These are conveniently expressed in terms of 
anticommutators at different times as  
\begin{eqnarray}\mlab{gfu2} 
i G^{>}_{ee}({\bf  k},t) &=& \sum_{r}  
\left[u^{r}_{{\bf k},1}\,   
\bar{u}^{r}_{{\bf k},1}  
\left\{\alpha^r_{{\bf k},e}(t),\alpha^{r\dag}_{{\bf k},e}\right\} 
e^{-i\omega_{k,1}t} +  \,
v^{r}_{-{\bf k},1}\,   
\bar{u}^{r}_{{\bf k},1}  
\left\{\beta^{r\dag}_{-{\bf k},e}(t),\alpha^{r\dag}_{{\bf k},e}
\right\}e^{i\omega_{k,1}t} \right],
\\ \mlab{gfu3}  
i G^{>}_{\mu e}({\bf  k},t) &=& \sum_{r}  
\left[u^{r}_{{\bf k},2}\, 
\bar{u}^{r}_{{\bf k},1}  
\,\left\{\alpha^r_{{\bf k},\mu}(t),\alpha^{r\dag}_{{\bf k},e}\right\}
e^{-i\omega_{k,2}t}  + 
\,v^{r}_{-{\bf k},2}\, \bar{u}^{r}_{{\bf k},1}  
\,\left\{\beta^{r\dag}_{-{\bf k},\mu}(t),  
\alpha^{r\dag}_{{\bf k},e}\right\}e^{i\omega_{k,2}t} \right] .
\end{eqnarray}
Here and in the following 
$\alpha^{r\dag}_{{\bf k},e}$ stands for $\alpha^{r\dag}_{{\bf k},e}(0)$.
We now have four distinct transition amplitudes, given by anticommutators
of flavor operators at different times:
\begin{eqnarray} \nonumber  
{\cal P}^r_{ee} ({\bf  k},t)  
 &\equiv& i \,u^{r \dag}_{{\bf k},1}{ e}^{i\omega_{k,1}t}\,  
 G^>_{ee}({\bf  k},t)\,\gamma^0 u^r_{{\bf k},1} =   
\left\{\alpha^r_{{\bf k},e}(t),\alpha^{r\dag}_{{\bf k},e} \right\}  
\\ \mlab{pee3}  
 &=&  \cos^{2}\!\theta\,  
 + \sin^2\!\theta\,\left[ |U_{\bf k}|^{2} { e}^{-i  
 (\omega_{k,2}-\omega_{k,1}) t}  
     + |V_{\bf k}|^{2} e^{i(\omega_{k,2}+\omega_{k,1}) t}\right],  
\\ \nonumber 
\\ \nonumber  
{\cal P}^r_{\bar{e}e} ({\bf  k},t)  
 &\equiv&i \,v^{r \dag}_{-{\bf k},1}{ e}^{-i\omega_{k,1}t}\,  
 G^>_{ee}({\bf  k},t)\,\gamma^0 u^r_{{\bf k},1} =  
  \left\{\beta^{r\dag}_{-{\bf k},e}(t),\alpha^{r\dag}_{{\bf k},e} \right\}  
\\ \mlab{pee4}  
 &=&  \epsilon^r\,|U_{\bf k}| |V_{\bf k}|\,\sin^{2}\!\theta\,  
 \left[  e^{i (\omega_{k,2}-\omega_{k,1})t}\;  
-\; e^{-i (\omega_{k,2}+\omega_{k,1})t} \right] , 
\\ \nonumber 
\\ \nonumber  
{\cal P}^r_{\mu e}({\bf  k},t)&\equiv& i \,u^{r   
\dag}_{{\bf k},2}e^{i\omega_{k,2}t}\,  
 G^>_{\mu e}({\bf  k},t)\,\gamma^0 u^r_{{\bf k},1} =   
\left\{\alpha^r_{{\bf k},\mu}(t),\alpha^{r\dag}_{{\bf k},e} \right\}   
\\ \mlab{pee5}  
&=&\;|U_{\bf k}|\;\cos\!\theta\;\sin\!\theta \left[1\;-\;  
 e^{i (\omega_{k,2}-\omega_{k,1}) t}\right],  
\\ \nonumber 
\\ \nonumber  
{\cal P}^r_{\bar{\mu} e}({\bf  k},t)&\equiv&  
i \,v^{r \dag}_{-{\bf k},2}e^{-i\omega_{k,2}t}\,  
 G^>_{\mu e}({\bf  k},t)\,\gamma^0 u^r_{{\bf k},1} =   
\left\{\beta^{r\dag}_{-{\bf k},\mu}(t),\alpha^{r\dag}_{{\bf k},e} \right\}   
\\ \mlab{pee6}  
&=&\;\epsilon^r\;|V_{\bf k}|\;\cos\!\theta\;\sin\!\theta   
\left[1\; -\; e^{-i (\omega_{k,2}+\omega_{k,1})t} \right]  
\;.\end{eqnarray}  
All other anticommutators with $\alpha^{\dag}_e $  
vanish. 
The probability amplitude is now  
correctly normalized:  $lim_{t\rightarrow 0^+}  
{\cal P}_{ee}({\bf  k},t)=1$, and ${\cal P}_{\bar{e}e}$, 
${\cal P}_{\mu e}$, ${\cal P}_{\bar{\mu} e}$ go to zero in the same 
limit $t\rightarrow 0^+\,$. Moreover, 
\begin{equation}\mlab{cons}  
\left|{\cal P}^r_{ee}({\bf  k},t)\right|^2 +   
\left|{\cal P}^r_{\bar{e}e}({\bf  k},t)\right|^2 +  
\left|{\cal P}^r_{\mu e}({\bf  k},t)\right|^2 +  
\left|{\cal P}^r_{\bar{\mu}e}({\bf  k},t)\right|^2 =1  
\,,\end{equation}  
as the conservation of the total probability requires. We also note  
that these transition probabilities are independent of the spin 
orientation.  

In order to understand the above transition amplitudes, 
consider the flavor charge operators, defined as  
$Q_{e/\mu}\equiv \sum_{{\bf k},r}
(\alpha_{{\bf k},e/\mu}^{r \dag}\alpha_{{\bf k},e/\mu}^r -  
\beta^{r\dag}_{-{\bf k},e/\mu}\beta_{-{\bf k},e/\mu}^r)$.  
We then have: 
\begin{eqnarray} \mlab{charge1} 
&&\;_{e,\mu}\langle 0(t)|Q_e|  
0(t)\rangle_{e,\mu}\; =\;_{e,\mu}\langle 0(t)|Q_{\mu}|  
0(t)\rangle_{e,\mu} \; =\; 0 \, ,
\\ \mlab{charge2} 
&& \langle \nu_e(t)|Q_e| \nu_e(t)\rangle\; = \; 
 \left|\left 
\{\alpha^{r}_{{\bf k},e}(t),  
\alpha^{r \dag}_{{\bf k},e} \right\}\right|^{2} \;+  
\;\left|\left\{\beta_{{-\bf k},e}^{r \dag}(t), 
\alpha^{r \dag}_{{\bf k},e} \right\}\right|^{2}
\, , 
\\ \mlab{charge3} 
&&\langle \nu_e(t)|Q_{\mu}| \nu_e(t)\rangle\; = \; 
 \left|\left\{\alpha^{r}_{{\bf k},\mu}(t), 
\alpha^{r \dag}_{{\bf k},e} \right\}\right|^{2} \;+  
\;\left|\left\{\beta_{{-\bf k},\mu}^{r \dag}(t), 
\alpha^{r \dag}_{{\bf k},e} \right\}  
\right|^{2}  .
\end{eqnarray}  

Charge conservation is ensured at any time: 
$\langle \nu_e(t)|\left(Q_e\;+\;Q_{\mu}\right)| \nu_e(t)\rangle\; = \; 1$
and the oscillation  
formula readily follows as
\begin{eqnarray} \mlab{enumber}  
P_{\nu_e\rightarrow\nu_e}({\bf k},t)&=& \left|\left 
\{\alpha^{r}_{{\bf k},e}(t),  
\alpha^{r \dag}_{{\bf k},e} \right\}\right|^{2} \;+  
\;\left|\left\{\beta_{{-\bf k},e}^{r \dag}(t), 
\alpha^{r \dag}_{{\bf k},e} \right\}\right|^{2}  
\\ \nonumber 
&=& 1 - \sin^{2}( 2 \theta) \left[ |U_{{\bf k}}|^{2} \;   
\sin^{2} \left( \frac{\omega_{k,2} - \omega_{k,1}}{2} t \right)  
+|V_{{\bf k}}|^{2} \;  
\sin^{2} \left( \frac{\omega_{k,2} + \omega_{k,1}}{2} t \right) \right]
\, , 
\\ \nonumber {} 
\\ \mlab{munumber}  
P_{\nu_e\rightarrow\nu_\mu}({\bf k},t)&=&  
\left|\left\{\alpha^{r}_{{\bf k},\mu}(t), 
\alpha^{r \dag}_{{\bf k},e} \right\}\right|^{2} \;+  
\;\left|\left\{\beta_{{-\bf k},\mu}^{r \dag}(t), 
\alpha^{r \dag}_{{\bf k},e} \right\}  
\right|^{2}  
\\ \nonumber 
&=&  \sin^{2}( 2 \theta)\left[ |U_{{\bf k}}|^{2} 
\;  \sin^{2} \left( \frac{\omega_{k,2} - \omega_{k,1}}{2} t \right)   
+|V_{{\bf k}}|^{2} \;  
\sin^{2} \left( \frac{\omega_{k,2} + \omega_{k,1}}{2} t \right) \right]
\, .
\end{eqnarray}  

This result is exact.
There are two differences with respect to the usual formula for neutrino
oscillations: the amplitudes are energy dependent, and there is an
additional oscillating term.
For $|{\bf k}|\gg\sqrt{m_1m_2}$, 
$|U_{{\bf k}}|^{2}\rightarrow 1$ 
and  $|V_{{\bf k}}|^{2}\rightarrow 0$ 
and the traditional oscillation formula is
recovered. 
However, also in this case we have that the neutrino state remains a
coherent state, thus phenomenological implications of our analysis 
are possible also for relativistic neutrinos. 
Further work in this direction is in progress.


\vspace{0.5cm}


\noi{\bf 4. Berry Phase for Oscillating Neutrinos}

\vspace{0.2cm}

Here I report on preliminary results\cite{ber} about  the existence of a
topological (Berry) phase\cite{Berry} 
in the evolution of a mixed state, more
specifically in neutrino oscillations.

Let us consider an electron neutrino state in the usual (Pontecorvo)
approximation\cite{BP78}; at time $t$ it is given by
\begin{equation}\lab{ber1}
|\nu_{e}(t)\rangle \equiv e^{-i H t} |\nu_{e}(0)\rangle= 
e^{-i \omega_{1} t} \left(\cos\theta\;|\nu_{1}\rangle \;+\;
e^{-i (\omega_{2}-\omega_{1}) t}\; \sin\theta\; |\nu_{2}\rangle \;
\right)\,.
\end{equation}
The state $|\nu_{e}(t)\rangle$, apart from
a phase factor, returns to the initial state
$|\nu_{e}(0)\rangle$ after a
period $T= \frac{2\pi}{\omega_{2} - \omega_{1}}$ :

\begin{equation}\lab{ber2}
|\nu_{e}(T)\rangle = e^{i \phi} |\nu_{e}(0)\rangle 
\;\;\;\;\;\;\;\;\;\;\;\;,\;\;\;\;\;\;\;\;\;\;\;\;\;
\phi= - \frac{2\pi \omega_{1}}{\omega_{2} - \omega_{1}} \,.
\end{equation}

It has been shown\cite{AA87} that any state which has a
cyclic quantum evolution, can acquire a geometrical (Berry) phase factor
after a cycle. This means that the phase factor $\phi$ of
eq.(\ref{ber2}) contains in general two contributions, a {\em
dynamical} one and a {\em geometrical} one. 
Thus the task is that of separate these two contributions. Following the
procedure stated in ref.\cite{AA87}, we get:
\begin{eqnarray}\nonumber
\beta&=& \phi + \int_{0}^{T}
\;\langle \nu_{e}(t)|\;H\;|\nu_{e}(t)\rangle \,dt
\\ \mlab{ber3}
&=&- \frac{2\pi \omega_{1}}{\omega_{2} - \omega_{1}} +
\frac{2\pi}{\omega_{2} - \omega_{1}}(\omega_{1}\;\cos^2\theta +
\omega_{2}\;\sin^2\theta)\;= \; 2 \pi \sin^{2}\theta  \,.
\end{eqnarray}
We thus see that there is indeed a non-zero geometrical phase $\beta$,
 related to the mixing angle
$\theta$, and that it is independent from the neutrino energies
 $\omega_1$, $\omega_2$ and masses $m_1,m_2$. 

An alternative way for calculating the geometrical phase is the
following. Define the state
\begin{equation}
|\tilde{\nu_{e}}(t)\rangle \equiv e^{-i f(t)} |\nu_{e}(t)\rangle \, ,
\end{equation}
with$f(t)=-\omega_{1}t$ such that $f(T) -f(0) = \phi$. Then the Berry
phase is defined as:
\begin{equation}
\beta=\int_{0}^{T}\;\langle \tilde{\nu_{e}}(t)|\;i \frac{d}{dt}
|\tilde{\nu_{e}}(t)\rangle\,dt \,= \; 2 \pi \sin^{2}\theta\, ,
\end{equation}
which coincides with the result (\ref{ber3}).

The topological phase factor of eq.(\ref{ber3}) acts then as a
``counter'' of oscillations: after each period (oscillation length), the
neutrino state gets an additional $\beta=2 \pi \sin^{2}\theta$ in its
phase. In principle, it is possible to think to
(interference) experiments in which one could measure this
phase, in analogy to what  is done in other situations 
(see ref.\cite{exber} for example): this would give a direct measurement
of the mixing angle.

\vspace{0.5cm}

\noi{\bf 5. Discussion and Conclusions}

\vspace{0.2cm}

I have shown how a simple dynamical system, such as the
one describing neutrino oscillations (eq.(\ref{lag2})), can exhibit many
interesting features if a proper analysis is carried out in the context
of Quantum Field Theory (QFT).

This is a crucial point: indeed it is well known that in QFT there exist
many inequivalent representations of the field algebra\cite{Um2} (many
vacua), and this makes the difference with
Quantum Mechanics, where only one Hilbert space is admitted. 
This considerations are far to be academic: I have shown in
eqs.(\ref{enumber}) and (\ref{munumber}) that the condensate 
structure of the flavor vacuum has physical
consequences on the neutrino oscillation formula.

It is also important to stress the generality of the above analysis: a
similar situation (with the due changes) 
holds for the case of mixing of
boson fields\cite{tesi} and in this respect the work is in progress.

This is true also for the topological 
 phase associated to ``flavor''
oscillations, which is  not peculiar 
of fermion systems but is a general
feature of mixed states of the form (\ref{ber1}), so the 
result (\ref{ber3}) is valid for a boson system as well. 
The work on the Berry phase is in progress\cite{ber}, 
in particular in the
direction of extending the above result 
to the three flavors case and to
the full QFT neutrino state (cf. eq.(\ref{h1})).

\vspace{0.5cm}

\noi{\bf Acknowledgements}

\vspace{0.2cm}

I would like thank Prof. T.Kibble for discussions and encouragement.
I also thank the organizers of the school, Profs. A.Zichichi,
G.'t Hooft and  G.Veneziano, and Prof. M.Gourdin 
for the support received as a scholarship
and for the invitation to contribute to these Proceedings. 
This work has been partially supported by MURST, INFN and ESF.


\newpage 

\begin{thebibliography}{9999}

\bibitem{Sol}  
R.Mohapatra and P.Pal, {\it Massive Neutrinos in Physics and  
Astrophysics}, (World Scientific, Singapore, 1991); \\  
J.N.Bahcall, {\it Neutrino Astrophysics},   
(Cambridge Univ. Press, Cambridge, 1989).

\bibitem{Fri}
H.Fritzsch and  Z.Z.Xing, hep-ph/9808272;
H.Fritzsch and Z.Z.Xing, {\it Phys.Rev.} {\bf D57} (1998) 594;
{\it Phys.Lett.} {\bf B413} (1997) 396.

\bibitem{Kam}
Super-Kamiokande Collaboration (Y. Fukuda et al.). {\it
Phys.Rev.Lett.} {\bf 81} (1998) 1562;
Kamiokande Collaboration(S. Hatakeyama et al.), 
{\it Phys.Rev.Lett.} {\bf 81} (1998) 2016.

\bibitem{BP78}  
S.M.Bilenky and B.Pontecorvo, {\it Phys.Rep.} {\bf 41} (1978) 225.


\bibitem{Fuj}
T.Kaneko, Y.Ohnuki and K.Watanabe, {\it Prog. Theor.Phys.}
{\bf 30} (1963) 521;\\
K.Fujii, {\it Il Nuovo Cimento} {\bf 34} (1964) 722;\\
K.Fujii, C.Habe and T.Yabuki, hep-ph/9807266.

\bibitem{GKL92} 
C.Giunti, C.W.Kim, J.A.Lee and U.W.Lee, {\it Phys.   
Rev.}{\bf D48} (1993) 4310; \\  
J.Rich, {\it Phys. Rev.}{\bf D48} (1993) 4318; \\
E.Sassaroli, hep-ph/9710239; hep-ph/9609476.


\bib{Baop}
M.Blasone and G. Vitiello,  {\it Ann. Phys. (N.Y.)} {\bf 244} 
(1995) 283; Erratum, ibid. {\bf 249} (1996) 363; 
E.Alfinito, M.Blasone, A.Iorio and G.Vitiello,   
 {\it Phys.Lett.} { \bf B362} (1995) 91.

\bib{ber}
M.Blasone, P.A.Henning  and G.Vitiello, in preparation.

\bib{Per}
A.Perelomov, {\it Generalized Coherent States and Their 
Applications}, 
(Springer-Verlag, Berlin, 1986).

\bib{BCS}
J.Bardeen, L.Cooper and J.Schrieffer, {\it Phys. Rev.} 
{\bf 108} (1957) 1175. 

\bib{Bpre2}
M.Blasone, P.A.Henning  and G.Vitiello,
hep-th/9803157.

\bibitem{Berry}
M.V.Berry, {\it Proc.Roy.Soc.London} {\bf A 392} (1984) 45.

\bibitem{AA87}
Y.Aharonov and J.Anandan, {\it Phys.Rev.Lett.} {\bf 58} (1987) 1593. 

\bib{exber}
A.G.Wagh et al., {\it Phys.Rev.Lett.} {\bf 78} (1997) 755.

\bib{Um2}
C.Itzykson and J.B.Zuber, {\it Quantum Field Theory},
(McGraw-Hill, New York, 1980);
H.Umezawa,{\it Advanced Field Theory: Micro, Macro and Thermal 
Physics} (American Institute of Physics, 1993).

\bib{tesi}
M.Blasone, PhD Thesis, 
Salerno University, December 1996.



\end{thebibliography}
\end{document}